\documentstyle[epsfig,eqsecnum,aps]{revtex}
\begin{document}
\draft
\title{$\beta$-decay in neutron-deficient Hg, Pb, and Po isotopes}

\author{O. Moreno,$^1$ P. Sarriguren,$^1$ 
R. \'Alvarez-Rodr\'{\i}guez,$^1$ and E. Moya de Guerra$^2$}
\address{$^1$Instituto de Estructura de la Materia,
Consejo Superior de Investigaciones Cient\'{\i }ficas, \\
Serrano 123, E-28006 Madrid, Spain}
\address{$^2$Departamento de F\'{\i}sica At\'omica, Molecular
y Nuclear, Facultad de Ciencias F\'{\i}sicas, \\ 
Universidad Complutense de Madrid, E-28040 Madrid, Spain}

%\date{\today}
\maketitle

\begin{abstract}

The effect of nuclear deformation on the energy distributions of the
Gamow-Teller strength is studied in neutron-deficient Hg, Pb, and Po 
even isotopes. The theoretical framework is based on a self-consistent
deformed Skyrme Hartree-Fock mean field with pairing correlations
between like nucleons in BCS approximation and residual spin-isospin
interactions treated in the proton-neutron quasiparticle random phase
approximation. After a systematic study of the Gamow-Teller strength
distributions in the low excitation energy region, relevant for
$\beta^+$-decay, we have identified the best candidates to look for
deformation signatures in their $\beta^+$-decay patterns. $\beta^+$
half-lives and total Gamow-Teller strengths $B(GT^\pm)$ are analyzed
as well.

\end{abstract}


\section{Introduction}
The existence of spherical and deformed nuclear shapes coexisting in a
given nucleus at low excitation energies is nowadays a well established
feature characterizing many isotopes in the neutron-deficient Hg, Pb,
and Po region (see Ref. \cite{julin} and references therein). These
coexisting structures have been found experimentally by studying the
$\gamma$-rays in coincidence with the emitted $\alpha$-particles in the
$\alpha$-decay of the fusion products created in selected
fusion-evaporation reactions. As a matter of fact, the existence of at
least one low-lying excited $0^+$ state in all even Pb isotopes between
$A=184$ and $A=194$ has been experimentally observed at excitation
energies below 1 MeV \cite{julin,andreyev}. Similarly, Hg isotopes from
$A=180$ up to $A=190$ have been found to be oblate in their ground
states with prolate excited states in the 1 MeV range.

Within a mean field description of the nuclear structure, the presence
of several $0^+$ states at low energies is understood as due to the
coexistence of different collective nuclear shapes. The energies of the
different shape configurations are calculated using a nuclear potential
with the energies of the single particle orbitals depending on the
deformation. Calculations of the potential energy surface have become
more and more refined with time. Phenomenological mean fields and
Strutinsky method \cite{bengtsson}, are already able to predict the
existence of several competing minima in the deformation energy surface
of neutron-deficient Hg and Pb isotopes. Self-consistent mean field
calculations \cite{smirnova,niksic02} and calculations including
correlations beyond mean field \cite{libert,egido,bender04} confirm
these results. In particular, for Pb isotopes, all the approaches
analyzed in Ref. \cite{egido}, from mean field up to very sophisticated
angular momentum projected generator coordinate methods, provide the
same underlying basic picture of strong coexisting spherical and
deformed shapes. This justifies the use of mean field approaches as a
first estimate to a qualitative description of the energy minima and
to associate the $0^+$ states with coexisting energy minima in this
energy surface. 

The relative energies of the different shapes predicted by mean field
calculations in the neutron-deficient Hg-Pb region have been found to
be very sensitive to fine details in the calculations, specially to
pairing effects. This is true in both non-relativistic \cite{tajima1} 
and relativistic calculations \cite{niksic02,yoshida94,yoshida97}.
In particular, standard deformed relativistic mean field (RMF)
calculations \cite{yoshida94,lala99} do not reproduce the experimentally
observed spherical ground states in the neutron-deficient Pb isotopes.
In RMF, these isotopes are found to be deformed using standard forces
with constant pairing gaps \cite{yoshida94,lala99}. It should be
emphasized that the same RMF+BCS calculations that lead to good agreement
with experimental data in systematic studies of ground state properties
carried out over 1300 even-even isotopes \cite{lala99}, fail to account
for the spherical ground states in this neutron-deficient Pb region.
It has been also shown \cite{yoshida97} that using constant strengths
for the pairing interaction, which makes the gap parameters dependent on
deformation, produces spherical ground states in the Pb isotopes. Yet,
more recently \cite{niksic02}, an improved treatment of pairing was used
by means of a relativistic Hartree-Bogoliubov calculation with the
NL3 effective interaction and a finite range Gogny force to describe
the pairing properties. However, it produced again oblate ground states
in $^{188-194}$Pb, contradicting the experimental data. In order to
recover the spherical ground states, a new parametrization of the
effective interaction was proposed \cite{niksic02}.

In a recent work \cite{sarri05}, we have calculated the $\beta$-decay
properties of the neutron-deficient Pb isotopes using a deformed Skyrme
HF+BCS+QRPA approach. This approach was used extensively \cite{sarr}
to demonstrate that the $\beta$-decay properties of unstable nuclei may
depend on the nuclear shape of the decaying parent nucleus and to predict
to what extent the GT strengths may be used as fingerprints of the
nuclear shapes. Accurate measurements of the GT strength distributions
in Kr and Sr isotopes \cite{isolde} have supported the usefulness of
these studies. By analyzing the energy deformation curves corresponding
to those Pb isotopes we found that, although the relative energies of
the various minima are dependent on both the Skyrme and pairing forces,
the existence of shape isomers as well as the location of their 
equilibrium deformations are rather stable. We also found that the
Gamow-Teller (GT) strength distributions calculated at the various
equilibrium deformations exhibit specific features that can be used as
signatures of the shape isomers, and what is important, these features
basically remain unaltered against changes in the Skyrme and pairing
forces. Therefore, although an accurate calculation of the excitation
energies of the $0^+$ states is beyond the scope of the theoretical 
framework used here, our estimate of the $\beta^+$-decay patterns is
reliable and has been proved to be useful \cite{sarr,isolde}.

Motivated by the interest in this type of calculations, not only in Pb 
isotopes but also in the neighbor regions where shape coexistence is 
found as well, in this paper we extend the calculations of the GT
strength distributions to the neutron-deficient
$^{180,182,184,186,188,190,192}$Hg, $^{184,186,188,190,192,194}$Pb and
$^{198,200,202,204,206}$Po even isotopes, using the Skyrme force
SLy4  \cite{sly4}, which is one of the most recent and successful forces. 
The reason for this choice of isotopes is that they are the best
candidates for $\beta$-decay experimental studies. They are 
$\beta^+$-unstable isotopes, yet not dominated by other decay modes
such as $\alpha$-decay. In addition they have relative large half-lives
and large enough $Q_{EC}$ energies to make experiments feasible.
Nevertheless, we present the results obtained not only within the
$Q_{EC}$ window relevant for the $\beta^+/EC$-decay, which is being
considered as a real possibility to be measured at ISOLDE/CERN
\cite{algora}, but also the GT strength distributions for $GT^+$ and
$GT^-$ in the whole range of excitation energies up to 25 MeV that
could be extracted from charge exchange reactions in the near future
at the new experimental facilities involving radioactive isotope beams,
such as FAIR. Charge exchange reactions at small momentum transfer are
a powerful tool to study GT strength distributions when $\beta$-decay
is not energetically possible. For incident energies above 100 MeV the
isovector spin-flip component of the complex effective interaction is
dominant and the cross section at forward angles is proportional to the
GT strength \cite{tad}.

The paper is organized as follows. In Sec. II we briefly present the
main features of our theoretical framework. Sec. III contains our results
on the energy deformation curves and Gamow-Teller strength distributions
in the neutron-deficient Hg, Pb, and Po isotopes relevant for 
$\beta^+$-decay. We also discuss the half-lives, the summed GT strengths,
and the GT strength distributions in the whole range of excitation
energies. Sec. IV contains the main conclusions.

\section{Brief description of the formalism}

The theoretical formalism used to describe the GT strength distributions 
has been already shown elsewhere \cite{sarr}. Here we only summarize the
basic ingredients of the method and discuss our choice of interactions
and parameters. The method starts from a self-consistent deformed
Hartree-Fock mean field calculation with effective two-body
density-dependent Skyrme interactions including pairing correlations in
BCS approximation. In this paper we consider the force SLy4 \cite{sly4},
which has been specially designed to improve the isospin properties of
nuclei away from beta stability and therefore, it may be more adequate
to describe the neutron-deficient Hg, Pb, and Po under study in this
work. Comparison with other widely used Skyrme forces like Sk3 \cite{sk3}
and SG2 \cite{sg2} was done in the past \cite{sarri05,sarr} concluding
that the main characteristics of the $\beta$-decay patterns are not
significantly changed when using different forces.

In this work we have restricted our study to axially deformed nuclear shapes.
The single-particle wave functions are then expanded in terms of the
eigenstates of an axially symmetric harmonic oscillator in cylindrical
coordinates, using eleven major shells. 
This simple picture is validated from triaxial calculations
\cite{bengtsson,libert,bender04}, which show at most slight triaxiality 
($\gamma \lesssim 10^\circ $) only in very few cases. However, the triaxial
barriers are not in general sufficiently high to fully support the axial 
symmetry assumption and it would be interesting to consider these
effects in future works. 

The method also includes pairing between like nucleons in the BCS
approximation with fixed gap parameters for protons $\Delta _{\pi},$
and neutrons $\Delta _{\nu}$, which are determined phenomenologically
from the odd-even mass differences through a symmetric five term formula
involving the experimental binding energies \cite{audi}. When the
experimental masses are not known we use the expression
$\Delta = 12 A^{-1/2}$ MeV for protons and neutrons. Occupation
probabilities are determined at each Hartree-Fock iteration by solving
the corresponding gap and number equations. The energy curves as a
function of deformation are calculated performing constrained HF
calculations with quadrupole constraints \cite{constraint}, minimizing
the HF energy for each nuclear deformation. 

We add to the mean field a separable spin-isospin residual interaction
to describe GT transitions. The advantage of using separable forces is
that the QRPA energy eigenvalue problem is reduced to find the roots
of an algebraic equation. The residual interaction contains two parts, 
particle-hole ($ph$) and particle-particle ($pp$). The $ph$ part 

\begin{equation}
V^{ph}_{GT} = 2\chi ^{ph}_{GT} \sum_{K=0,\pm 1} (-1)^K \beta ^+_K 
\beta ^-_{-K}, \qquad 
\beta ^+_K = \sum_{\pi\nu } \left\langle \nu \left| \sigma _K \right|
\pi \right\rangle a^+_\nu a_\pi \, ,
\end{equation}
is responsible for the position and the general structure of the GT
resonance \cite{sarr,moller,homma}. The usual procedure to determine
the coupling strength $\chi ^{ph}_{GT}$ is to reproduce the energy
of the resonance \cite{homma}. In this work we use the same value of 
$\chi ^{ph}_{GT}$ for all the isotopes considered in this mass region.
The value is fixed by comparison of the calculations in $^{208}$Pb to
the experimental GT strength resonance observed at an energy of 19.2 MeV,
relative to the parent nucleus, from $(p,n)$ charge exchange reactions 
\cite{gaarde}. In this way, we reproduce the energy of the resonance
with $\chi ^{ph}_{GT}= 0.08$ MeV. 

The particle-particle residual interaction is a neutron-proton pairing
force in the $J^\pi=1^+$ coupling channel. We introduce this interaction
as a separable force in the usual way \cite{homma,hir}

\begin{equation}
V^{pp}_{GT} = -2\kappa ^{pp}_{GT} \sum_K (-1)^K P ^+_K P_{-K}, \qquad 
P ^+_K = \sum_{\pi\nu} \left\langle \pi \left| \left( \sigma_K\right)^+
\right|\nu \right\rangle  a^+_\nu a^+_{\bar{\pi}} \, .
\end{equation}
Once the value of the $ph$ spin-isospin residual force has been established,
the coupling strength $\kappa ^{pp}_{GT}$ is usually adjusted to reproduce
the measured half-lives \cite{homma,hir} because the peak of the GT
resonance is almost insensitive to the $pp$ force. In our case, as
explained in Ref. \cite{sarri05} we have found that the calculated
half-lives as a function of $\kappa ^{pp}_{GT}$ are very little dependent
on this parameter and we use $\kappa ^{pp}_{GT}=0.02$ MeV, which is
compatible with the parametrization of Ref. \cite{homma}, obtained from
a systematic fitting procedure over the nuclear chart.

For even-even nuclei, the Gamow-Teller strength $B(GT^{\pm})$ in the
laboratory frame for a transition 
$I_{i}^{\pi_i} K_i (0^+0)\rightarrow I_{f}^{\pi_f} K_f(1^+K)$
can be obtained as

\begin{equation}
B(GT^{\pm})= \sum_{M_i,M_f,\mu} \left| \left< I_fM_f \left| 
\beta ^\pm _\mu \right| I_i M_i \right> \right|^2= \left\{ \delta_{K_f,0}
\left< \phi_{K_f} \left|  \beta ^\pm _0 \right| \phi_0\right> ^2 +2
\delta_{K_f,1} \left< \phi_{K_f} \left|  \beta ^\pm _1 \right| 
\phi_0\right> ^2 \right\} \, ,
\label{streven}
\end{equation}
in units of $g_A^2/4\pi$ and in terms of the intrinsic amplitudes
calculated in QRPA\cite{sarr}. To obtain this expression we have used
the initial and final states in the laboratory frame expressed in terms
of the intrinsic states $|\phi_K >$, using the Bohr-Mottelson factorization
\cite{bm}.

The $\beta$-decay half-life is obtained by summing up all the allowed 
transition probabilities, weighed with phase space factors, up to 
states in the daughter nucleus with excitation energies lying within 
the $Q_{EC}$-window,

\begin{equation}
T_{1/2}^{-1}=\frac{A^2}{D}\sum_{\omega }f\left( Z,\omega \right) B(GT)\, ,
 \label{t12}
\end{equation}
where $D=6200$~s. We include the standard effective factor

\begin{equation}
A^{2}=\left[ \left( g_{A}/g_{V}\right) _{\rm eff}\right] ^{2}=\left[
0.77\left( g_{A}/g_{V}\right) _{\rm free}\right] ^{2} \, . \label{quen}
\end{equation}

In this work we use experimental $Q_{EC}$ values because as it was 
discussed in Ref. \cite{sarri05} the $Q_{EC}$-values calculated from the
binding energies of parent and daughter nuclei are quite similar for the
various shapes and are close to the corresponding experimental values.
In $\beta^+/EC$ decay, $f\left( Z,\omega \right) $ consists of two parts,
positron emission and electron capture. We have computed them 
numerically for each value of the energy.
The Fermi integrals $f^{\beta^\pm}\left( Z,\omega \right) $ are given by

\begin{equation}
f^{\beta^\pm} (Z, W_0) = \int^{W_0}_1 p W (W_0 - W)^2 \lambda^\pm(Z,W) 
{\rm d}W\, , 
\end{equation}
with

\begin{equation}
\lambda^\pm(Z,W) = 2(1+\gamma) (2pR)^{-2(1-\gamma)} e^{\mp\pi y}
\frac{|\Gamma (\gamma+iy)|^2}{[\Gamma (2\gamma+1)]^2}\, ,
\end{equation}
where $\gamma=\sqrt{1-(\alpha Z)^2}$ ; $y=\alpha ZW/p$ ; $\alpha$ is the fine 
structure constant and $R$ the nuclear radius. $W$ is the total energy of the 
$\beta$ particle, $W_0$ is the total energy available in $m_e c^2$ units, and
$p=\sqrt{W^2 -1}$ is the momentum in $m_e c$ units. In the numerical calculation,
we have included screening and finite size effects as explained in 
Ref. \cite{gove}. The electron capture rates $f^{EC}$ have also been included
following Ref. \cite{gove}:

\begin{equation}
f^{EC}=\frac{\pi}{2} \sum_{x} q_x^2 g_x^2B_x
\end{equation}
where $x$ denotes the atomic subshell from which the electron is captured,
$q$ is the neutrino energy, $g$ is the radial component of the bound state
electron wave function at the nucleus, and $B$ stands for other exchange and
overlap corrections \cite{gove}.

\section{Results and discussion}

In this section we present the results obtained for the $\beta$-decay
patterns of the neutron-deficient  $^{180-192}$Hg, $^{184-194}$Pb and
$^{198- 206}$Po even isotopes. First, we discuss the energy surfaces
and shape coexistence expected in these isotopes. Then, we present the
results obtained for the Gamow-Teller strength distributions with
special attention to their dependence on the nuclear shape and discuss
their relevance as signatures of deformation to be explored experimentally.
Finally, we discuss the half-lives and the summed GT strengths both within
the $Q_{EC}$ and in the whole energy range.

\subsection{Equilibrium deformations}
  
We show in Fig. 1 the HF energies calculated with the Skyrme force SLy4
and pairing correlations treated in the fixed gap approach. We show the
results for Hg (left), Pb (middle), and Po (right) isotopes as a function
of the quadrupole deformation parameter calculated as
$\beta=\sqrt{\pi/5}Q_p/(Z<r^2>)$, which is defined in terms of the proton
quadrupole moment $Q_p$ and charge r.m.s. radius $<r^2>$. The curves have
been scaled to the energy of their ground states and have been shifted by
1 MeV from one isotope to the next one, starting from the lightest one.

In the case of Hg isotopes we get prolate and oblate minima in all the 
isotopes from $A=180$ up to $A=192$. The ground state is prolate for
$^{180}$Hg and $^{182}$Hg and oblate for $^{184-192}$Hg isotopes.
Comparable results were also obtained in RMF calculations \cite{lala99}, 
where oblate ground states where obtained at $\beta \sim -0.2$ with
prolate minima close in energy at $\beta \sim 0.13$. This is in
qualitative agreement with experiment, where it is observed that a weakly 
oblate ground-state band is crossed in the vicinity of $A=188$ by a
deformed band associated with a prolate energy minimum. The prolate states
minimize their energies for $A=182$, although they still lie above the
ground state. We get qualitative agreement although the details of the
energy minima depend much on the effective Skyrme force and pairing
treatment used in the calculations. In the case of Pb isotopes we obtain
a prolate ground state for $^{184}$Pb  and oblate ground states for
$^{186-194}$Pb isotopes. As we can see in the figure, prolate and oblate
minima are always present from $A=184$ to $A=190$. Spherical local
minima can be clearly seen also in $A=184$ and shallow profiles are found
in other cases. For Po isotopes we find oblate ground states in all the
isotopes considered $^{198-206}$Po. We also find prolate solutions, which
are more prominent in $A=200,202,204$ isotopes. 

In general, we observe that both oblate and prolate deformed minima occur
at smaller deformation as the number of neutrons increases, approaching
the $N=128$ closed shell. This is true in Hg, Pb, and Po isotopes and the
effect is more pronounced in the prolate shapes. The oblate minima are
rather stable at $\beta = -0.2$ in Hg and Pb isotopes and at about 
$\beta = -0.1$ in Po isotopes. 

We get similar qualitative results for the Hg, Pb, Po isotopes considered
when other Skyrme forces are used. More specifically, we obtain the same
patterns of coexistence with minima at about the same deformations although
the relative energies may change from one force to another.
In the case of Hg and Pb isotopes the correspondence of the minima obtained
with different forces is almost perfect. In the case of Po isotopes we find
the deformation of the minima with SG2 and SLy4 almost coincident. Sk3 shows
only one oblate minimum that corresponds quite well with the oblate minimum
obtained with the other forces, but in the prolate region Sk3 shows only a
small distortion of the curve without developing a minimum. In the case of
$^{206}$Po, Sk3 and SG2 exhibit only a flat minimum centered at the spherical
shape.

In general, our results agree with experiment in the sense that we find
different coexisting shapes in this mass region, but in some cases they
are at variance with experiment in the predicted shape of the ground state.
This is especially true with respect to Pb isotopes. This discrepancy, also
found in different theoretical frameworks \cite{niksic02,tajima1}, was
discussed in \cite{sarri05}, where we demonstrated on some examples the
sensitivity of the energy curves to fine details of the two-body
interactions, showing that the deformation minima remain at about the same
deformation values but their relative energies may change considerably.
      
\subsection{Gamow-Teller strength distributions}

In this subsection we study the energy distribution of the Gamow-Teller 
strengths calculated at the equilibrium shapes that minimize the energy
of the nucleus.  

In Fig. 2 we show the GT strength distributions in ($g_A^2/4\pi$) units
as a function of the excitation energy in the daughter nucleus for the
Hg isotopes under study. The results correspond to QRPA with the force
SLy4, fixed pairing gaps obtained from experimental masses and residual
separable interactions $ph$ and $pp$ with coupling strengths given by
$\chi ^{ph}_{GT}=0.08$ MeV and $\kappa ^{pp}_{GT}=0.02$ MeV, respectively. 
The left panels contain the discrete strengths for both prolate
(upward) and oblate (downward) equilibrium deformations. The dashed arrows
indicate the experimental $Q_{EC}$ values of the decay. 
We have used the same energy scale to appreciate better the differences
as we increase the number of neutrons. The scale of energies is 6 MeV,
enough to include all the $Q_{EC}$ energies, whose maximum value is
$Q_{EC}=5.352$ MeV in $^{180}$Hg. The right panels
contain the accumulated GT strength for prolate (thick lines) and oblate 
(thin lines) deformations. In this case the energy axis is extended up to
the $Q_{EC}$ value in order to observe directly the allowed energy window 
and the total GT strength expected in the decay.
Although such kind of accumulated plots contain in principle the same 
information as the pure spectrum, it offers a more global and appealing
view, which is sometimes preferable.
One should also notice that in figures 2-7 the quenching factor
(see Eq.(\ref{quen})) is not included and therefore, one should consider
a reduction of this strength by a factor of about two when comparing to
experiment.

From Fig. 2 one can see that prolate and oblate nuclear shapes give rise
to GT strength distributions whose patterns can be clearly distinguished
from each other. The oblate strength appears in all cases very fragmented
over the whole $Q$-window, which induces a straight increasing of the
strength in the accumulated plot. On the other hand, the distributions
in the prolate cases show a concentration of the GT strength in a single
strong peak located at very low excitation energy (below 0.7 MeV in all
the isotopes). This single excitation carries practically the whole
strength contained in the $Q$-window in the $A=190,192$ isotopes and
about half of it in $A=180-188$.
The accumulated plots are then characterized by the successive appearance
of large vertical steps followed by large flat regions, in contrast to 
the oblate case. With the help of the accumulated plots we can also observe
immediately that although the total strength contained within the $Q_{EC}$
energy is quite similar for oblate and prolate shapes, the distribution 
of this strength along the energy axis is very different.

Fig. 3 contains the same information as Fig. 2 but for Pb isotopes. In
this case we also include results for spherical shapes in such a way
that the left panels contain the results from prolate shapes (upper
plots), oblate shapes (middle plots) and spherical shapes (lower plots),
except in the $A=194$ isotope where we only have two shapes. Similarly,
the right panels contain three lines, thick line for prolate, thin for
oblate, and dashed for spherical shapes. As in the case of Hg isotopes,
the signature of the prolate shapes is a strong isolated peak at low
excitation energies (there is a second strong peak in $A=188$). This 
makes again the accumulated plots to appear as big steps. The oblate 
distributions show a very fragmented structure, starting at higher
energies than the strong prolate peak. This produces an accumulated
pattern which increases in a steady way, starting at higher energy than
the prolate pattern. In the spherical case, the strength is practically
collected in a single peak, which corresponds to the 
$\pi h_{11/2} \rightarrow \nu h_{9/2}$ transition. This excitation
carries less strength than the prolate case and occurs in all the
isotopes except in $A=184$ at an energy higher than the 
prolate peaks. It is also worth noticing that oblate and prolate shapes
generate practically the same total GT strength within the $Q_{EC}$ window
although the internal distribution of this strength is in general very
different. The total strength in the spherical case is always clearly
below the prolate or oblate strength, except in the isotope $A=194$, where
the small strength and $Q_{EC}$ makes it not very relevant.

Fig. 4 shows the same results as in Fig. 2 for Po isotopes. In the cases
of $A=198,200$ we find in the prolate case below $Q_{EC}$ a strong single
peak carrying most of the strength, while again the oblate shapes produce
a very fragmented distribution. In the other cases, $A=202,206$,
the strength below  $Q_{EC}$ is insignificant and could not tell much
to distinguish the shapes.

In general, we observe that the GT strength distributions corresponding
to prolate shapes in both Hg and Pb isotopes are characterized by the
appearance of a strong excitation below 1 MeV that collects a large amount
of the total strength contained in the whole region below the $Q_{EC}$
energy. On the other hand, the GT strength distributions corresponding
to oblate shapes are always very fragmented and are extended over the
whole $Q_{EC}$ window. In the case of the Pb isotopes this strength is
located at excitation energies higher than the prolate peak. For Pb
isotopes the GT strength distributions corresponding to spherical shapes
appear as single peaks carrying all the strength below $Q_{EC}$ and also
at higher energies than the prolate peak. These features are in agreement
with the conclusions in Ref. \cite{sarri05} using the force Sk3.
The most noticeable difference with respect to the results obtained from
Sk3 force \cite{sarri05} appears in the spherical case, where the GT
strength obtained with Sk3 is displaced to higher energies as compared
to the strength obtained with SLy4. The origin of the shift in the
excitation energy can be traced back to the predicted energies for the
$h_{9/2}$ and $h_{11/2}$ spherical shells. The main contribution to the
GT strength at low energies comes from the transition connecting the
almost fully occupied $\pi h_{11/2}$ shell with the partially unoccupied
$\nu h_{9/2}$ shell, and the relative position of the two shells
determines the GT excitation energy. The effect of deformation is to allow
for multiple transitions fragmenting the strength and smoothing the
differences caused by the forces.

In Figs. 5-7 we show the $B(GT^{\pm})$ strength distributions in the Hg, Pb
and Po isotopes. We plot continuous distributions resulting from a folding
procedure using $\Gamma=1$ MeV width gaussians on the discrete spectrum.
The energy range of excitation energies in the daughter nucleus is extended
up to 20 MeV in order to cover all the strength. We find the strength beyond
20 MeV to be insignificant, except for Po isotopes where we extend the scale
up to 25 MeV. The interest of these figures, not limited by the $Q_{EC}$
energies, is to show the whole strength distributions that could be explored
from charge exchange reactions such as $(n,p)$ and $(p,n)$ or similar
reactions. The $B(GT^-)$ strength on the right panels is much stronger than
the corresponding $B(GT^+)$ strength on the left panels, as it should be
according to the Ikeda sum rule
$\Sigma_{\omega} [B(GT^-)(\omega) -B(GT^+)(\omega)] = 3(N-Z)$,
which is fulfilled in all our calculations up to a small few percent
discrepancy. Ikeda sum rule ranges from $3(N-Z)=60$ in $^{180}$Hg up to
$3(N-Z)=114$ in $^{206}$Po. We can see in the last three columns of Tables
1, 2, and 3 the total sums of the $B(GT^+)$, $B(GT^-)$, and the percentage
of the Ikeda sum rule fulfilled in the calculation for Hg, Pb, and Po
isotopes, respectively.

In Fig. 5, the profiles of the $B(GT^-)$ strength show a first bump, which
for the lightest isotopes considered is a double bump, moving from 4 MeV in
$A=180$ up to 7 MeV in $A=192$. We can also see a big resonance at higher
energies, that moves from 11 MeV in $A=180$ up to 14 MeV in $A=192$.
In any case, the profiles obtained from oblate and prolate shapes are quite
similar and then, it will be very difficult to use this information to 
distinguish between one shape or another. In the case of the $B(GT^+)$
strength, there is a resonance at low energy below 2 MeV, which is rather
constant in all the isotopes. As we have seen in more detail in Fig. 2, it
is in this region below $Q_{EC}$ where the shape dependence could be
exploited because prolate and oblate shapes produce distinguishable
distributions of the GT strength.

Fig. 6 is similar to Fig. 5 for Pb isotopes. Again we see in the profiles
of the $B(GT^-)$ strength on the right panels a small bump between 5 and
6 MeV and a resonance centered at 12-13 MeV. The nuclear shapes can hardly
be identified looking at the strength distributions. The $B(GT^+)$
strengths show also a resonance at low energies, which corresponds to the
analysis made in Fig. 3. Fig. 7 for Po isotopes shows once more the small
bump in the $B(GT^-)$  distribution at 6-7 MeV and the big resonance at
12-14 MeV with similar characteristics in all the isotopes and shapes.
The $B(GT^+)$ strength shows in this case a resonance at high energy
around 20 MeV and the low energy distributions that were studied in Fig.4.

\subsection{Half-lives and summed strengths}

It is also interesting to calculate integral magnitudes from the GT strength
distributions that characterize them in a single quantity. This is the case
of the total half-lives and total GT strength contained in the energy range
available in the $\beta^+$-decay. These magnitudes depend differently on
the GT strength distribution, one of them is simply a sum and the other one
is a weighted sum (see Eq.(\ref{t12})). While the total GT strength involved
in the $\beta^+$-decays are not measured yet, there is experimental
information on the half-lives\cite{audi} that allows to contrast our
calculations and to check that there is no substantial disagreement with
experiment. We can see the results in Tables 1, 2, and 3 for Hg, Pb, and Po
isotopes, respectively. The first column in Tables 1-3 contains the
experimental $Q_{EC}$ energies \cite{audi}. Since the $\beta^+ /EC$ decay
mode competes in this mass region with $\alpha$-decay, we show in the second
column of the tables the percentage of the total decay assigned to
$\beta^ +$ \cite{audi}. The third column contains the total experimental
half-lives \cite{audi} and within brackets the corresponding $\beta^ +$
half-lives extracted from the percentage in the second column. The next
columns contain theoretical results calculated at the equilibrium shapes
obtained with the force SLy4. We can see first the $\beta^+/EC$ half-lives
calculated with quenching factors (\ref{quen}) and experimental $Q_{EC}$
values, as discussed in Ref. \cite{sarri05}. Next, we can see the $B(GT^+)$
strength summed up to the $Q_{EC}$ energies, the total $B(GT^-)$ and
$B(GT^+)$ sums and the percentage of the Ikeda sum rule fulfilled.

In Table 1 one can see that the half-lives obtained for Hg isotopes agree
in general with experiment within a factor of 2. 
Only the most stable isotope $A=192$ deviates considerably from this
agreement, but this case is not very relevant because of the small energy
window allowed for the decay $Q_{EC}=0.765$ MeV and because of the small
strength involved, which makes it meaningless.
A factor of 2 is perfectly acceptable in this type of calculations taking
into account the uncertainties coming from the forces used, the level
of approximations, the quenching factors and the $Q_{EC}$ energies.
For instance, reasonable quenching factors could change the half-lives 
within a $25\%$ effect and varying $Q_{EC}$ values within 1 MeV range could
change the half-lives by a factor of two. Another source of uncertainty
in the half-lives is induced by the determination of the pairing gaps
from the experimental binding energies. Through variations of $20\%$
below and above the adopted proton and neutron pairing gap values,
we obtain half-lives which in general do not differ by more than $20\%$
from the original results.
We also find that in the spherical cases the half-lives are more sensitive
to these uncertainties than in the deformed cases.
Nevertheless, as we have discussed, this does not rule out the conclusions
extracted based on the patterns shown by the GT strength distributions.
We also include in the table the strength contained below the $Q_{EC}$
energy that could be measured in $\beta$-decay
experiments \cite{algora}. We should keep in mind that in contrast to the
calculation of the half-lives, we do not include quenching factors in
the summed strengths and thus, a reduction of about $50\%$ of the values
written in the tables are expected when comparing to experiment.
The sums over the whole energy region of the $B(GT^+)$ and $B(GT^-)$ 
strengths show that the total $B(GT^-)$ is much stronger than $B(GT^+)$
as it should happen in order to fulfill Ikeda sum rule, as it was
discussed above. In the case of Hg isotopes we can see that the half-lives
from prolate deformations are in general larger than those from oblate
ones with the already mentioned exception of $A=192$.
The summed strengths are practically the same no matter what the
deformation is.

The same general comments can be applied also to Table 2 for Pb isotopes. 
We get in general half-lives that agree with experiment within a factor
2-3 . This is also true when we consider other Skyrme forces. In particular,
the half-lives obtained with the forces Sk3 and SG2 are very similar to the
corresponding SLy4 ones shown in the tables. The only disagreement appears
in the spherical cases. We find in the spherical case differences
up to factors of 4-5 in the half-lives obtained from various forces.
As we have already discussed above, this is due to the different predicted
energies of the spherical shell gaps.
If we compare the half-lives from different deformations, we can see that
the half-lives from the spherical shapes are larger than those
corresponding to the deformed shapes, and the prolate shapes give always
lower half-lives than the oblate ones. In the case of SLy4, the half-lives
from the spherical shapes agree with experiment at the same qualitative
level as the half-lives from deformed shapes. This is at variance with
what we found in  Ref. \cite{sarri05} using Sk3 and SG2 forces, in the
sense that there, we only found agreement with experimental half-lives
with deformed shapes. The strengths summed up to $Q_{EC}$ are very close
to each other for oblate and prolate deformations and are much larger
than the corresponding strength for the spherical case.

In Table 3 we have the results for Po isotopes. In this case the total GT
strength involved in the decay is very small in all the cases and
therefore the half-lives are large and not very meaningful. Only in the
two more unstable isotopes $A=198,200$ this comparison makes sense and
we can see that again the half-lives agree with experiment within a factor
of two. The oblate half-lives are larger than the corresponding prolate
ones.

Although the deformation does not show up in the total GT strengths,
it does in the internal distribution of the GT strength, specially in
the $B(GT^+)$ at low excitation energy where the $\beta^+$-decay is
sensitive. The only purpose of showing half-lives and summed strengths
is to contrast the validity of the calculation against some measured
quantities in the case of half-lives or to predict the expected total
strength coming into play in the other case.

\section{Summary and Conclusions}

We have studied the energy distribution of the Gamow-Teller strength in 
the neutron-deficient $^{180-192}$Hg, $^{184-194}$Pb, and $^{198-206}$Po
isotopes. We use a deformed pnQRPA formalism with spin-isospin $ph$ and
$pp$ separable residual interactions. The $ph$ coupling strength is
fixed to reproduce the energy of the GT resonance in $^{208}$Pb.
The $pp$ coupling strength is taken from existing global parametrizations,
although we find our results to be quite insensitive to this value.
The quasiparticle mean field is generated from a deformed HF approach
with the two-body Skyrme effective interaction SLy4. It includes pairing
correlations in BCS approximation, using fixed gap parameters extracted
from the experimental masses. The equilibrium deformation is derived
self-consistently within the HF procedure. 

An analysis of the energy deformation curves shows that oblate and prolate
coexisting shapes are expected in the neutron-deficient Hg and Po isotopes.
In the case of Pb isotopes we also get in some cases a spherical minimum
giving rise to a triple shape coexistence.
We should remark again that the relative energies of the minima in these
calculations are very sensitive to the Skyrme force and type of pairing
used and therefore the predicted shapes for the ground states can be
altered depending on the choice of these forces.
However, for a given isotope, the equilibrium deformations at which the
minima occur are hardly shifted.

The GT strength distributions calculated at the equilibrium shapes
show that the effect of the deformation is much stronger than the effects
coming from the Skyrme or pairing forces used. As a general rule, we find
that the GT strength distributions calculated from prolate shapes are
characterized by strong single peaks at low excitation energies carrying
a sizable amount of the total GT strength involved in the $\beta^+$-decay.
The small remaining GT strength is scattered over the energy interval up
to $Q_{EC}$. On the other hand, the signature of oblate shapes is
a distribution spread all over the available $Q$-window energy.
In the case of Pb isotopes, the GT strength distributions calculated from
spherical shapes are characterized by single peaks that contain all the
GT strength, which are in general located at higher energies than the
prolate peaks. The detailed analysis performed allows us to identify
good candidates to explore experimentally the signatures of deformation
on the GT strength distributions. We find that the most interesting
isotopes are $^{184-188}$Hg, $^{188-192}$Pb and $^{198,200}$Po.

We have also investigated the GT strength distributions in the whole
range of excitation energies without the limitations imposed by the
$\beta^+$-decay. This analysis is worth as a prediction for the cross
sections to be expected in charge exchange reactions. 

The detailed analysis at low excitation energies in the $B(GT^+)$
distribution shows that the $\beta^+$-decay process is the most sensitive
to the nuclear shape.

The calculated total strength involved in the  $\beta^+$-decay shows
little dependence on the nuclear shape, except in the case of the 
spherical shapes in Pb isotopes, where this strength is much smaller.
The calculated $\beta^+ /EC$ half-lives agree with experiment within
a factor of 2 or 3. In general we can conclude that neither the half-lives
nor the summed strengths are good observables to study deformation effects,
but the strength distributions are.

\vskip 1cm 

\begin{center}
{\Large \bf Acknowledgments} 
\end{center}
We are grateful to A. Algora and B. Rubio for useful discussions.
This work was supported by Ministerio de Educaci\'on y Ciencia (Spain)
under contract number FIS2005-00640. O.M. thanks Ministerio
de Educaci\'on y Ciencia (Spain) for financial support. R.A.-R. thanks
I3P Programme (CSIC, Spain) for financial support. We also acknowledge
participation in the EXL-EURONS European Collaboration (RI3-506065).

\newpage

\begin{center}

\begin{table}[t]
\caption{ Half-lives and $B(GT^{\pm})$ strengths in Hg isotopes. The
table contains experimental $Q_{EC}$ values [MeV], percentage of the
$\beta^+/EC$ involved in the total decay, total experimental half-lives
[s] \protect\cite{audi} and within brackets, the $\beta^+/EC$ experimental
half-lives extracted from the percentage. Then we find theoretical 
results obtained with the force SLy4:  half-lives [s],
$B(GT^+)$ strength $[g_A^2/(4\pi)]$ summed up to $Q_{EC}$ energies, 
total $B(GT^+)$ and $B(GT^-)$ strengths contained in the whole
energy range considered $(E_{ex}<30)$ MeV, and percentage of the
Ikeda sum rule fulfilled in our calculations.}
\label{table.1}
\begin{tabular}{rcccccccc}\cr
 isotope  & $Q_{EC}$ & \% $\beta^+$ & $T_{1/2,exp}^{\rm total}$   
 $(\beta^+)$ & $T_{1/2,SLy4}^{\beta^+/EC}$ & 
 $\Sigma _{Q_{EC}} B(GT^+)$ & $\Sigma B(GT^+)$ & $\Sigma B(GT^-)$ & \% Ikeda
\cr \cr
\hline
\cr
$^{180}$Hg, obl & 5.352 & 52 & 2.6 (4.9) & 3.0 & 3.10 & 4.63 &
  64.20 & 99.28 \cr 
 prol & &&& 4.6 & 2.73 & 4.43 & 63.99 & 99.27 \cr \cr
$^{182}$Hg, obl & 4.725 & 86.2 & 10.8 (12.6) & 5.6 & 2.49 & 4.27 &
  69.62 & 99.02 \cr
 prol & &&& 7.5 & 2.28 & 4.18 & 69.51 & 98.98 \cr \cr
$^{184}$Hg, obl & 3.970 & 98.89 & 30.6 (30.9) & 17.5 & 1.71 & 3.06 &
  74.46 & 99.17 \cr
 prol & &&& 17.0 & 1.46 & 3.08 & 74.50 & 99.19 \cr \cr
$^{186}$Hg, obl & 3.176 & 100 & 82.8  (82.8) & 47.2 & 1.22 & 2.57 &
  79.85 & 99.08 \cr
 prol & &&& 68.4 & 0.89 & 2.46 & 79.76 & 99.10 \cr \cr
$^{188}$Hg, obl & 2.099 & 100 & 195 (195) & 185.1 & 0.78 & 2.51 &
  85.64 & 98.96 \cr
 prol & &&& 218.9 & 0.74 & 2.59 & 85.72 & 98.96 \cr \cr
$^{190}$Hg, obl & 1.511 & 100 & 1200 (1200) & 506.9 & 0.63 & 2.10 &
  90.84 & 98.60 \cr
 prol & &&& 493.8 & 0.47 & 2.27 & 91.35 & 98.98 \cr \cr
$^{192}$Hg, obl & 0.765 & 100 & 17460 (17460) & 5200 & 0.14 & 1.71 &
  96.55 & 98.79 \cr
 prol & &&& 1970 & 0.27 & 1.80 & 96.94 & 99.10 \cr
\end{tabular}
\end{table}

\begin{table}[t]
\caption{ Same as in Table 1 for Pb isotopes.}
\begin{tabular}{rcccccccc}\cr
 isotope  & $Q_{EC}$ & \% ($\beta^+/EC$) & $T_{1/2,exp}^{\rm total}$
 $(T_{1/2,exp}^{\beta^+/EC})$ & $T_{1/2,SLy4}^{\beta^+/EC}$ & 
 $\Sigma _{Q_{EC}} B(GT^+)$ & $\Sigma B(GT^+)$ & $\Sigma B(GT^-)$ & \%
\cr \cr
\hline
\cr
$^{184}$Pb, sph & 5.84 & 20 & 0.49 (2.45) & 5.6 & 0.85 & 1.95 &
  61.65 & 99.50 \cr 
 obl & &&& 5.6 & 2.31 & 3.61 & 63.31 & 99.50 \cr 
 prol & &&& 3.3 & 2.55 & 4.01 & 63.70 & 99.48 \cr \cr
$^{186}$Pb, sph & 5.51 & 60 & 4.82  (8.03) & 16.3 & 0.54 & 1.65 &
  67.28 & 99.44 \cr
 obl & &&& 10.7 & 1.73 & 2.91 & 68.58 & 99.50 \cr 
 prol & &&& 4.8 & 1.91 & 3.34 & 68.94 & 99.39 \cr \cr
$^{188}$Pb, sph & 4.53 & 90.7 & 25.5  (28.1) & 28.0 & 0.76 & 2.11 &
  73.76 & 99.51 \cr
 obl & &&& 20.3 & 1.35 & 3.05 & 74.72 & 99.54 \cr 
 prol & &&& 16.1 & 1.46 & 3.16 & 74.78 & 99.47 \cr \cr
$^{190}$Pb, sph & 3.92 & 99.6 & 71 (71) & 80.0 & 0.54 & 1.92 &
  79.48 & 99.44 \cr
 obl & &&& 41.7 & 0.99 & 2.67 & 80.26 & 99.47 \cr
 prol & &&& 26.4 & 0.99 & 2.89 & 80.32 & 99.27 \cr \cr
$^{192}$Pb, sph & 3.32 & 100 & 210 (210) & 251.7 & 0.41 & 1.90 &
  85.57 & 99.61 \cr
 obl & &&& 93.4 & 0.70 & 2.44 & 86.12 & 99.62 \cr
 prol & &&& 45.6 & 0.79 & 2.70 & 86.26 & 99.48 \cr \cr
$^{194}$Pb, sph & 2.62 & 100 & 720 (720) & 655.8 & 0.62 & 2.02 &
  90.70 & 98.53 \cr
 obl & &&& 482.5 & 0.56 & 2.32 & 90.99 & 98.52 \cr 
\end{tabular}
\end{table}

\begin{table}[t]
\caption{ Same as in Table 1 for Po isotopes.}
\begin{tabular}{rcccccccc}\cr
 isotope  & $Q_{EC}$ & \% ($\beta^+/EC$) & $T_{1/2,exp}^{\rm total}$   
 $(T_{1/2,exp}^{\beta^+/EC})$ & $T_{1/2,SLy4}^{\beta^+/EC}$ & 
 $\Sigma _{Q_{EC}} B(GT^+)$ & $\Sigma B(GT^+)$ & $\Sigma B(GT^-)$ & \%
\cr \cr
\hline
\cr
$^{198}$Po, obl & 3.900 & 43 & 106 (247) & 342 & 0.387 & 2.23 &
  91.13 & 98.78 \cr 
 prol & &&& 160 & 0.333 & 2.24 & 91.25 & 98.90 \cr \cr
$^{200}$Po, obl & 3.416 & 88.9 & 690 (776) & 1390 & 0.085 & 1.82 &
  96.64 & 98.77 \cr
 prol & &&& 1066 & 0.189 & 1.82 & 96.76 & 98.90 \cr \cr
$^{202}$Po, obl & 2.809 & 98.08 & 2682 (2735) & 8799 & 0.029 & 1.42 &
  102.9 & 99.46 \cr
 prol & &&& 5341 & 0.015 & 1.41 & 102.88 & 99.48 \cr \cr
$^{204}$Po, obl & 2.334 & 99.34 & 12708 (12729) & 29788 & 0.013 & 1.32 &
  108.88 & 99.59 \cr
 prol & &&& 12167 & 0.018 & 1.31 & 108.90 & 99.62 \cr \cr
$^{206}$Po, obl & 1.846 & 94.55 & 760320 (804145) & 256650 & 0.0037 & 1.25 &
  114.78 & 99.59 \cr
 prol & &&& 55352 & 0.0046 & 1.24 & 114.72 & 99.54 \cr
\end{tabular}
\end{table}

\newpage
\begin{figure}[t]
\epsfig{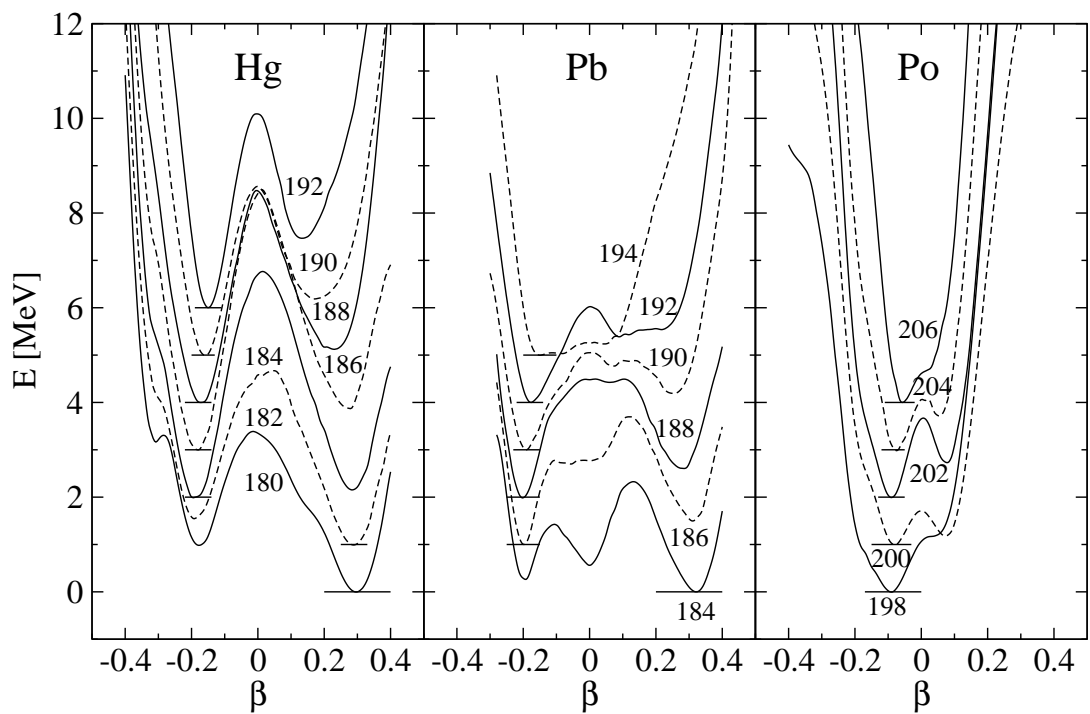}
\vskip 1cm
\caption{HF energy scaled to the ground state energy (see text) obtained
from constrained HF+BCS calculations with the Skyrme force SLy4
and fixed pairing gap parameters as a function of the quadrupole 
deformation $\beta$.
Left:  $^{180-192}$Hg isotopes, middle: $^{184-194}$Pb isotopes, and 
right: $^{198-206}$Po isotopes.}
\end{figure}

\newpage

\begin{figure}[t]
\epsfig{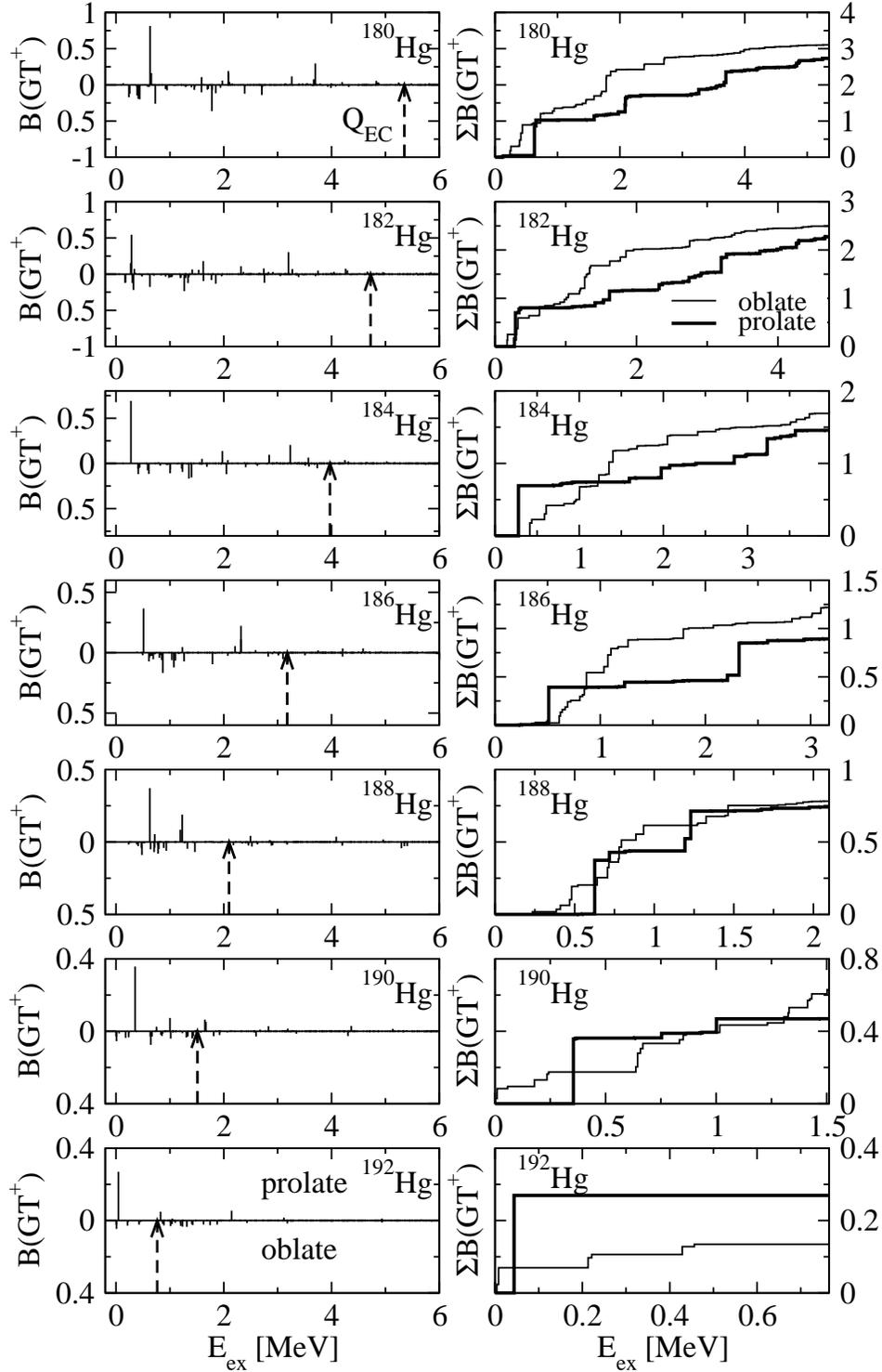}
\vskip 1cm
\caption{Left: Gamow-Teller strength distributions $[g_A^2/(4\pi)]$
in Hg isotopes for prolate (upward) and oblate (downward) shapes. The 
experimental $Q_{EC}$ energies are shown by dashed arrows. Right: 
Accumulated Gamow-Teller strength for prolate (thick lines) and oblate 
(thin lines) shapes plotted up to the $Q_{EC}$ energies.
Results are obtained from SLy4 force with fixed gap parameters.}
\end{figure}

\newpage

\begin{figure}[t]
\epsfig{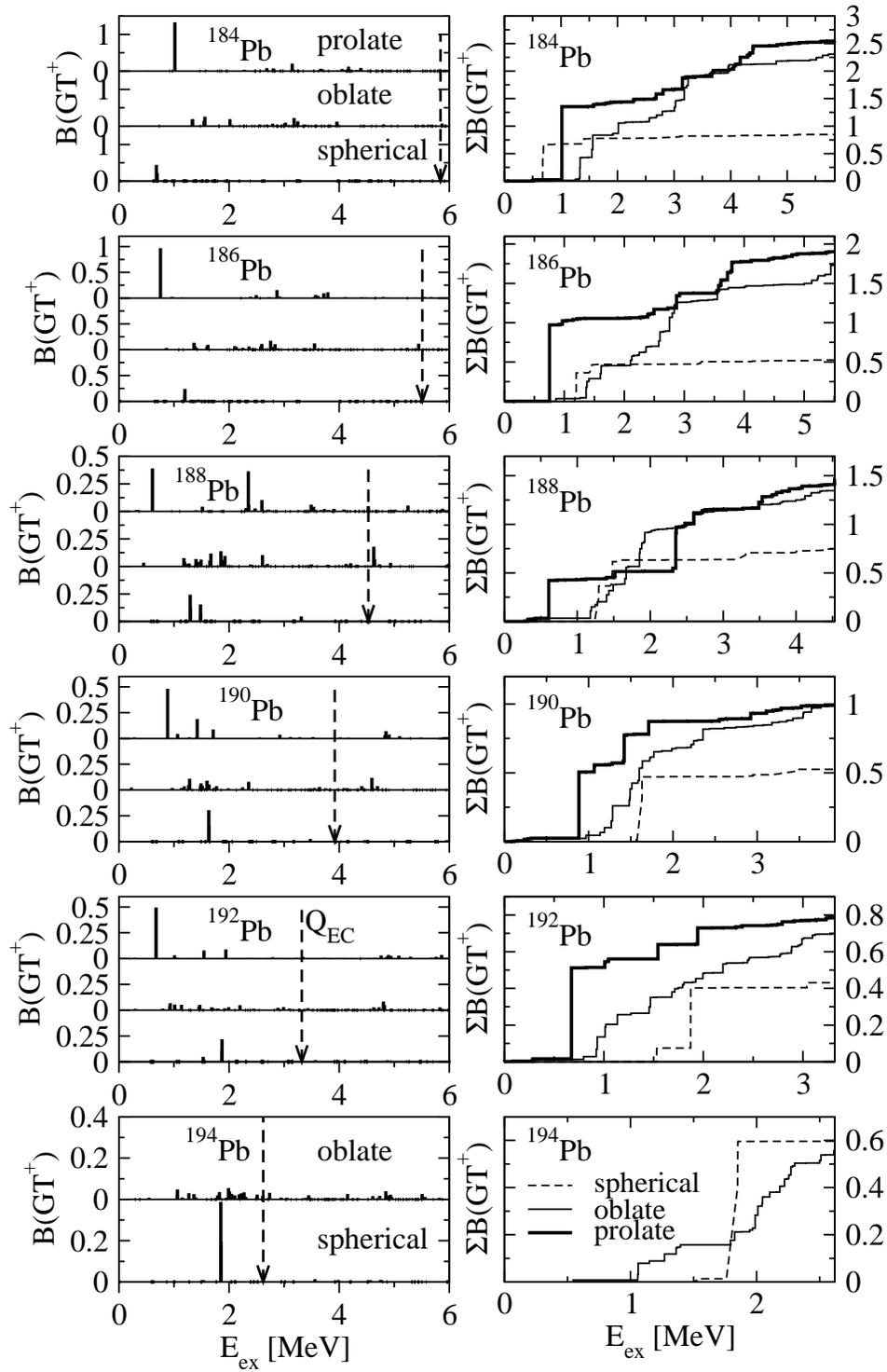}
\vskip 1cm
\caption{Same as in Fig. 2 for Pb isotopes. In this case we also include results
from spherical shapes.}
\end{figure}

\newpage

\begin{figure}[t]
\epsfig{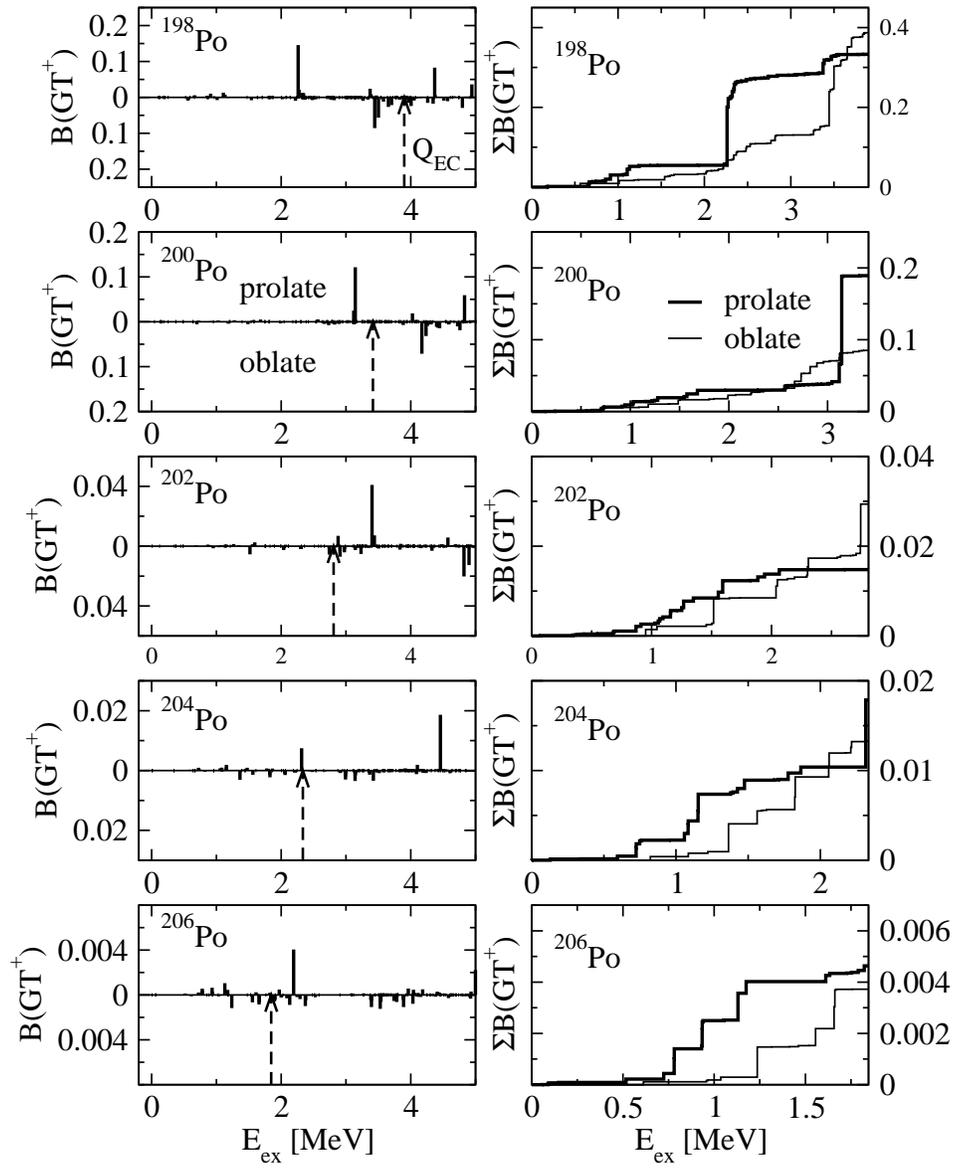}
\vskip 1cm
\caption{Same as in Fig. 2 for Po isotopes.}
\end{figure}

\newpage

\begin{figure}[t]
\epsfig{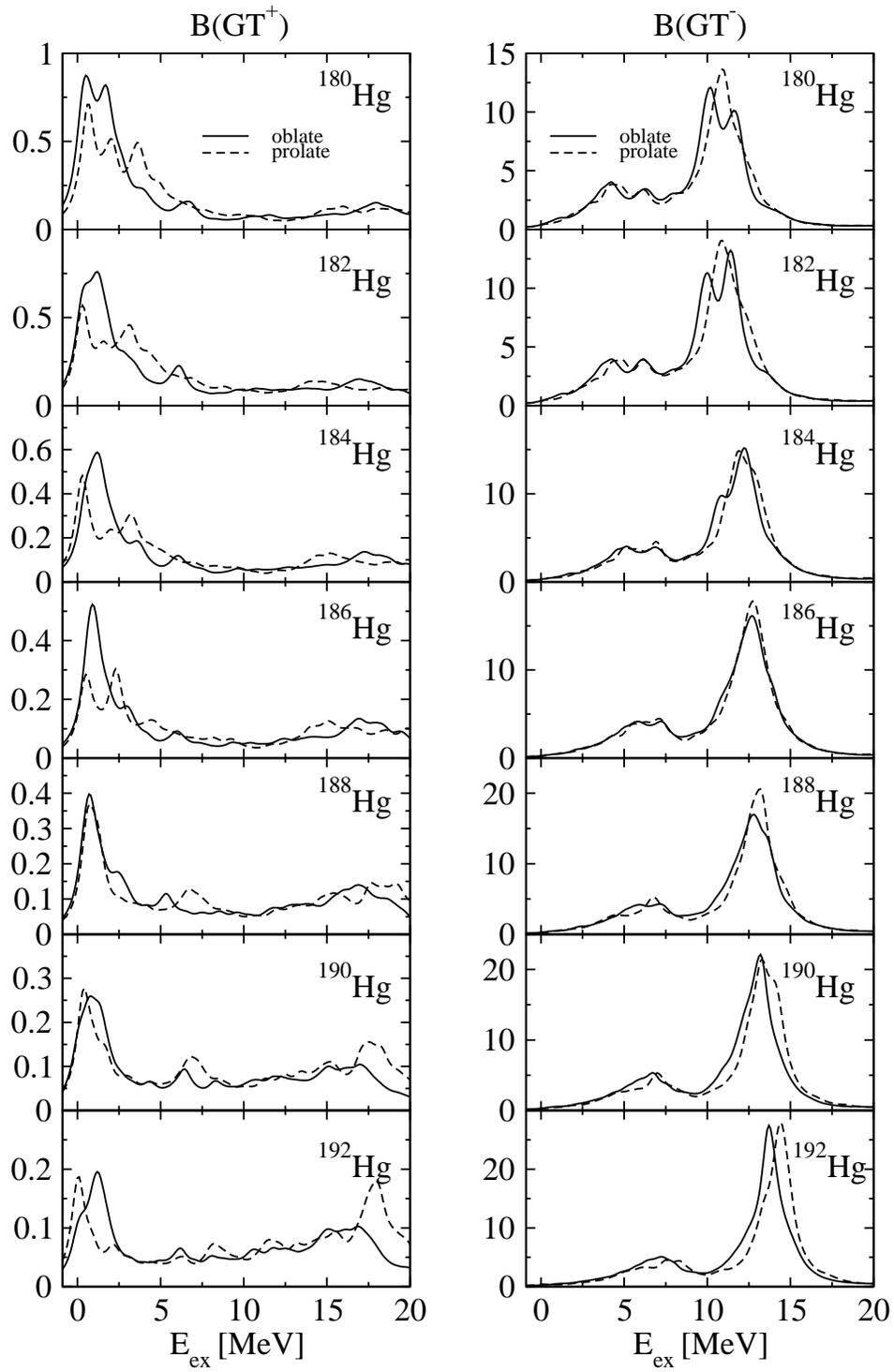}
\vskip 1cm
\caption{Folded Gamow-Teller strength distributions 
$B(GT^{\pm})\; [g_A^2/(4\pi)]$ for the various shapes of the Hg 
isotopes in the range of excitation energies up to 20 MeV.}
\end{figure}

\newpage

\begin{figure}[t]
\epsfig{file=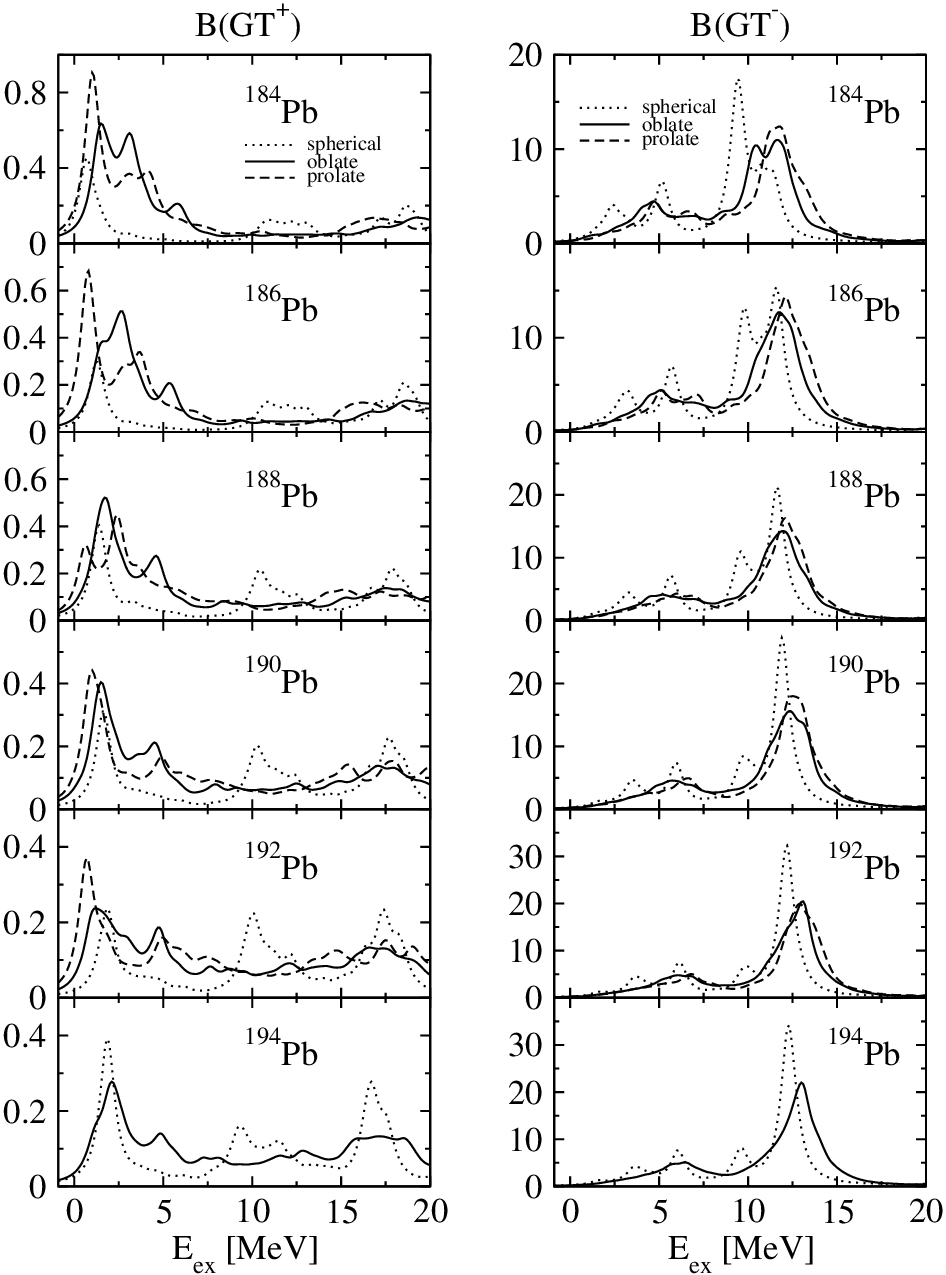,width=0.7\textwidth}
\vskip 1cm
\caption{Same as in Fig. 5 for Pb isotopes.}
\end{figure}

\newpage

\begin{figure}[t]
\epsfig{file=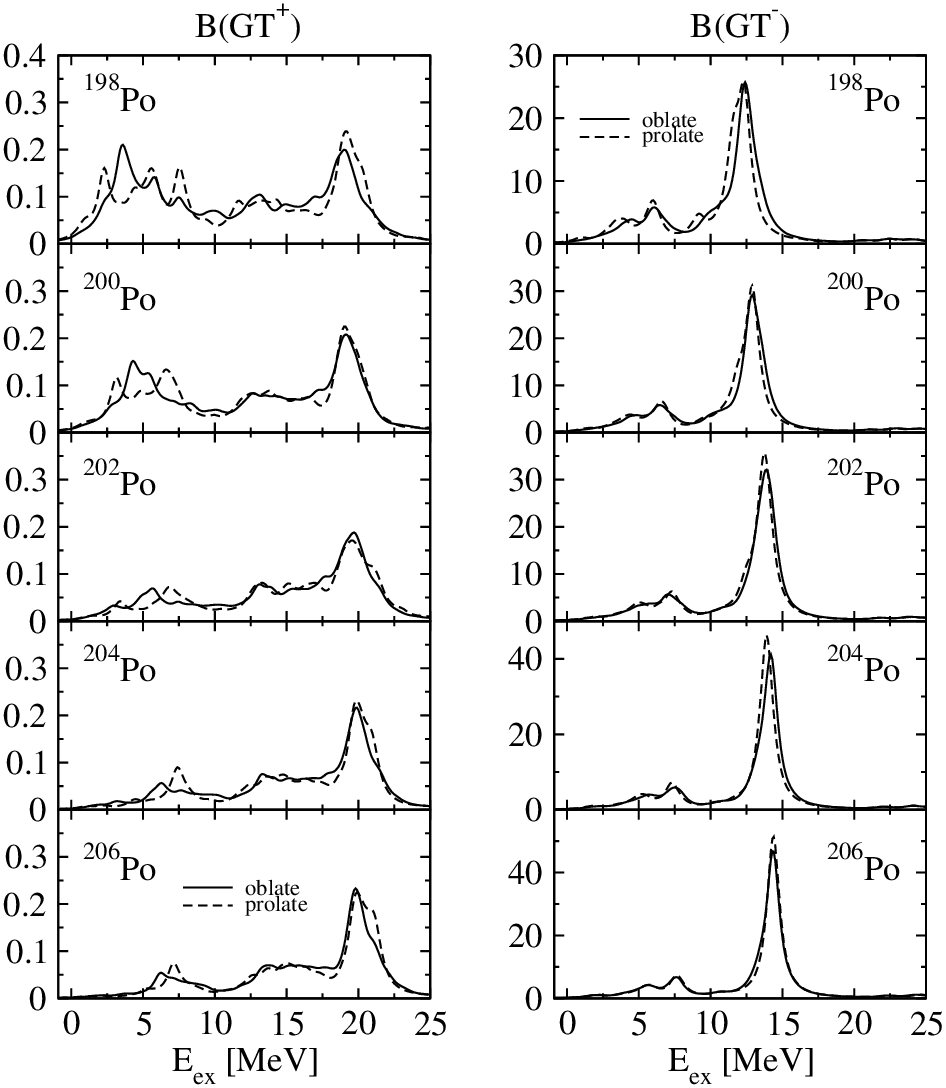,width=0.7\textwidth}
\vskip 1cm
\caption{Same as in Fig. 5 for Po isotopes.}
\end{figure}

\end{center}


\newpage

\begin{thebibliography}{00}

\bibitem{julin} R. Julin, K. Helariutta and M. Muikku, J. Phys. G.: Nucl.
Part. Phys. {\bf 27}, R109 (2001).

\bibitem{andreyev} A.N. Andreyev {\em et al.}, Nature {\bf 405}, 430 (2000).

\bibitem{bengtsson} R. Bengtsson and W. Nazarewicz, Z. Phys. A {\bf 334}, 
269 (1989); W. Nazarewicz, Phys. Lett. B {\bf 305}, 195 (1993).

\bibitem{smirnova} N.A. Smirnova, P.-H. Heenen and G. Neyens, Phys. Lett. B
{\bf 569}, 151 (2003).

\bibitem{niksic02} T. Niksic, D. Vretenar, P. Ring and G.A. Lalazissis, Phys.
Rev. C {\bf 65}, 054320 (2002).

\bibitem{libert} J. Libert, M. Girod and J.-P. Delaroche, Phys. Rev. C
{\bf 60}, 054301 (1999).

\bibitem{egido} J.L. Egido, L.M. Robledo and R.R. Rodr\'{\i}guez-Guzm\'an,
Phys. Rev. Lett. {\bf 93}, 082502 (2004); R.R. Rodr\'{\i}guez-Guzm\'an, 
J.L. Egido and L.M. Robledo, Phys. Rev. C {\bf 69}, 054319 (2004).

\bibitem{bender04} M. Bender, P. Bonche, T. Duguet and P.-H. Heenen, Phys.
Rev. C {\bf 69}, 064303 (2004).

\bibitem{tajima1} N. Tajima, H. Flocard, P. Bonche, J. Dobaczewski and 
P.-H. Heenen, Nucl. Phys. A {\bf 551}, 409 (1993).

\bibitem{yoshida94} S. Yoshida, S.K. Patra, N. Takigawa and C.R. Praharaj,
Phys. Rev. C {\bf 50}, 1398 (1994).

\bibitem{yoshida97} S. Yoshida and N. Takigawa, Phys. Rev. C {\bf 55}, 
1255 (1997).

\bibitem{lala99} G.A. Lalazissis, S. Raman, and P. Ring, Atomic Data and
Nuclear data Tables {\bf 71}, 1 (1999).

\bibitem{sarri05} P. Sarriguren, O. Moreno, R. \'Alvarez-Rodr\'{\i}guez, and 
E. Moya de Guerra, Phys. rev. C {\bf 72}, 054317 (2005).

\bibitem{sarr}  P. Sarriguren, E. Moya de Guerra, A. Escuderos and A.C.
Carrizo, Nucl. Phys. A {\bf 635}, 55 (1998); P. Sarriguren, E. Moya de 
Guerra and A. Escuderos, Nucl. Phys. A {\bf 658}, 13 (1999); Nucl. Phys. 
A {\bf 691},  631 (2001); Phys. Rev. C {\bf 64}, 064306 (2001). 

\bibitem{isolde} E. Poirier {\it et al.}, Phys. Rev. C {\bf 69}, 034307 
(2004); E. N\'acher {\it et al.}, Phys. Rev. Lett. {\bf 92}, 232501 (2004).

\bibitem{sly4} A. Chabanat, P. Bonche, P. Haensel, J. Meyer, and R.
Schaeffer, Nucl. Phys. A {\bf 635}, 231 (1998).

\bibitem{algora} A. Algora, B. Rubio and W. Gelletly, private communication.

\bibitem{tad} T.N. Taddeucci {\it et al.}, Nucl. Phys. A {\bf 469}, 125 (1987).

\bibitem{sk3}  M. Beiner, H. Flocard, N. Van Giai and P. Quentin, Nucl.
Phys. A {\bf 238}, 29 (1975).

\bibitem{sg2}  N. Van Giai and H. Sagawa, Phys. Lett. B {\bf 106}, 
379 (1981).

\bibitem{audi} G. Audi, O. Bersillon, J. Blachot and A.H. Wapstra,
Nucl. Phys. A {\bf 729}, 3 (2003).

\bibitem{constraint}  H. Flocard, P. Quentin, A.K. Kerman and D. Vautherin,
Nucl. Phys. A {\bf 203}, 433 (1973).

\bibitem{moller} J. Krumlinde and P. Moeller, Nucl. Phys. A {\bf 417}, 
419 (1984); P. Moeller and J. Randrup, Nucl. Phys. A {\bf 514}, 1 (1990).

\bibitem{homma} H. Homma, E. Bender, M. Hirsch, K. Muto, H.V.
Klapdor-Kleingrothaus and T. Oda, Phys. Rev. C {\bf 54}, 2972 (1996).

\bibitem{gaarde} C. Gaarde et al. Nucl. Phys. A {\bf 369}, 258 (1981);
{\it ibid.} {\bf 396}, 127c (1983).

\bibitem{hir}
M. Hirsch, A. Staudt, K. Muto and H.V.
Klapdor-Kleingrothaus, Nucl. Phys. {\bf A535}, 62 (1991);
K. Muto, E. Bender, T. Oda and H.V. Klapdor-Kleingrothaus,
Z. Phys. A {\bf 341}, 407 (1992).

\bibitem{bm} A. Bohr and B. Mottelson, {\em Nuclear Structure}, (Benjamin,
New York 1975).

\bibitem{gove} N.B. Gove and M.J. Martin, Nucl. Data Tables {\bf 10}, 205 
(1971).


\end{thebibliography}
\end{document}